\documentclass[%
 reprint,
 amsmath,amssymb,
 aps,
]{revtex4-2}
\usepackage{graphicx}
\usepackage{dcolumn}
\usepackage{bm}
\usepackage[colorlinks=true,linkcolor=blue]{hyperref}
\hypersetup{
	colorlinks,%
	citecolor=blue,%
	filecolor=blue,%
	linkcolor=blue,%
	urlcolor=blue
}
\DeclareGraphicsExtensions{.png,.pdf,.jpg,.tif}
\begin{document}
\title{Entangled two-photon absorption detection through a Hong-Ou-Mandel interferometer}
\author{Freiman Triana-Arango$^{1,}$}%
\email{freiman@cio.mx}
\author{Gabriel Ramos-Ortiz$^{1,}$}%
\email{garamos@cio.mx}
\author{Roberto Ramírez-Alarcón$^{1,}$}%
\email{roberto.ramirez@cio.mx}
\affiliation{$^{1}$Centro de Investigaciones en Óptica AC, Apartado Postal 37150, León, Gto, México}
\begin{abstract}
Recently, different experimental methods to investigate the entangled two-photon absorption (ETPA) process in a variety of materials have been reported. The present work reports on a different approach on which the ETPA process is detected based on the changes induced on the Hong-Ou-Mandel (HOM) interferogram of two photons produced by a spontaneous parametric down conversion (SPDC) Type-II process. Using an organic solution of the laser dye Rhodamine B as a model, it is demonstrated that ETPA at $800nm$ can be monitored by the changes in the temporal-width and visibility of the HOM interferogram produced by the photons at a beam splitter after interacting with the sample. Additionally, we present a detail model in which the sample is modeled as a filtering function which fullfills the energy conservation conditions required by ETPA, allowing to explain the experimental observations with excellent agreement. We believe that this work represents a new perspective to study the ETPA process, by using an ultra-sensitive quantum interference technique and a detailed mathematical model of the process.
\end{abstract}
\maketitle
\section{\label{sec:Introduction}Introduction}
The non-linear optical phenomenon of two-photon absorption (TPA) is of current great interest, as it finds various scientific and technological applications, such as laser scanning (multiphoton) microscopy, microengraving, photodynamic therapy, etc \cite{WDenk, Wei07, Aparicio-Ixta2016}. Typically, TPA is achieved in third order materials with large TPA cross-sections $(\delta_{c})$ through the use of pulsed lasers delivering high density of random photons. However, from recent theoretical \cite{Fei,FrankSchlawin} and experimental work \cite{Dayan_exp,Goodson3}, the interest for implementing TPA with correlated (entangled) photons, namely entangled two-photon absorption (ETPA) process, has increased notoriously.
	
The advantage of the ETPA process compared to the classical TPA effect is that the former contributes to the total rate of absorbed photons $(R_{TPA})$ with a linear dependence on the photon excitation flux $(\phi)$, while the dependence is quadratic in the case of classical light \cite{FrankSchlawin,Perina,Dayan_teo}: $R_{TPA}=\sigma_{e}\phi+\delta_{c}\phi^{2}$, where $\sigma_{e}$ is entangled two-photon absorption cross-section. Notoriously, there can be $~32$ orders of magnitude difference between $\sigma_{e}$ \cite{Goodson3,Tabakaev,Goodson4} and $\delta_{c}$ \cite{Makarov08,Xu96,Sperber}, $(\sigma_{e}\sim10^{-18}[cm^{2}/molecule],~\delta_{c}\sim10^{-50}[cm^{4}s/molecule])$, so in principle it is possible to achieve TPA in the low flux regime by illuminating the sample with correlated photon pairs produced by quantum processes such as spontaneous parametric down conversion (SPDC) \cite{SPDC_Adel}. 

The calculated values of $\sigma_{e}$ reported in the literature for molecules used as models lies in a wide range from $10^{-22}$ to $10^{-18}[cm^{2}/molecule]$ \cite{JuanVillabona1, Tabakaev, KristenM}, however there is a recent debate about the real magnitude of this parameter or even the possibility to experimentally observe the effect \cite{KristenM,Samuel_Alfred}. 

Different experimental configurations have been employed to detect and quantify the ETPA activity in different materials, based on measuring the transmittance \cite{Goodson3,Goodson4,Goodson1,JuanVillabona1,Samuel_Alfred} and fluorescence exhibited by the sample upon excitation with SPDC down-converted photon pairs \cite{Dayan_exp,Goodson2,KristenM,TiemoLandes}. Although it has been pointed out that the ETPA quantification in transmittance experiments could be influenced by artifacts \cite{KristenM}, it is expected that $\sigma_{e}$ lays within the aforementioned range. In such experiments the difference in transmittance from a solution of the molecule under test and the solvent alone is recorded, as a function of the excitation intensity, with the experimental condition of the temporal delay ($\delta t$) between the down-converted photons set to zero. In many of these works, the zero delay condition is assured by implementing a HOM interferometer before the photons interact with the sample. Recently, the HOM interference have been used to analyze the changes in one of the photons of the down-converted pair by using a Mach-Zehnder interferometer to analyze a additional time delay introduced by a sample \cite{Audrey_Goodson}, however the temporal-width and visibility changes in the HOM dip, once the down-converted photons interacted with the sample, has not been consider as a measurement device to study the ETPA interaction. 
\begin{figure*}[htbp!] 
	\includegraphics[width=\textwidth]{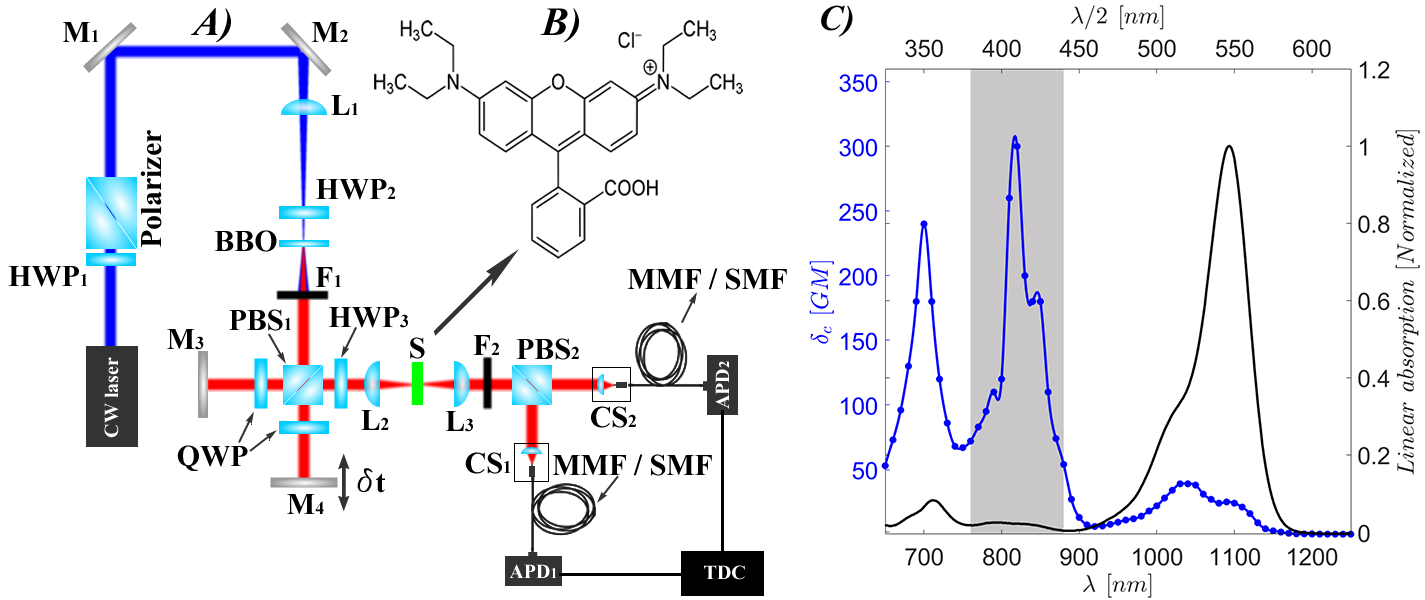}
	\caption{A) Experimental configuration for the transmittance ETPA experiments designed to use the HOM interferometer as a sensing device. B) Rhodamine B molecule. C) Rhodamine B TPA cross-section (blue dots) and Rhodamine B linear absorption spectrum (black solid curve) as a function of wavelength. The gray rectangle corresponds to the the excitation region in our ETPA transmittance experiment. The experimental data represented as blue dots was taken from \cite{Makarov08}.} 
	\label{setup2}
\end{figure*}
In this context, the present work proposes a transmittance experiment in which the ETPA interaction in a sample is identified and quantified by the changes induced in the visibility and temporal-width of the HOM interferogram produced by two indistinguishable photons, generated in a SPDC Type-II process, after they both interact with the sample. 
\section{\label{sec:Experimental-section}Experiment}
The experimental setup is shown in Fig.(\ref{setup2}A). A CW laser (Crystalaser DL-405-100), centered at $403nm$ and with a $\sim 1nm$ bandwidth, is focused by lens L1 (focal length $f_1=500mm$) into a BBO crystal to produce collinear cross-polarized frequency-degenerate Type-II SPDC photons pairs around $806nm$. In order to optimize the fulfill of the phase-matching conditions, which maximize the SPDC emission process, it is necessary to align the polarization of the pump beam with the plane defined by the crystal's optical axis. We do so by rotating the zero order half wave plate $HWP_{2}$, which also works as a control of the density of down-converted photons emitted by the crystal. Prior being focused to enter the crystal, the laser pump power is controlled by a half-plate-wave ($HWP_{1}$) and a Glan-Thompson polarizer ($Polarizer$). After the crystal we use the filter element $F_1$, composed of a longpass filter (Thorlbas FELH0500) and a bandpass filter (Thorlbas FBH800-40), in order to eliminate the residual pump. Then, the filtered down-converted photons propagates through a Michelson interferometer (starting at $PBS_{1}$) which introduces a controllable temporal delay ($\delta t$) between them. 

Additionally to control the frequency (bandapass filter) and temporal (delay) indistinguishability, the final requisite to obtain an optimal HOM interference pattern at $PBS_{2}$ is to eliminate the polarization distiguishability, proper of photon pairs produced in a Type-II SPDC process. To do so, a half-waveplate ($HWP_{3}$) is introduced to rotate $45^{\circ}$ the horizontal and vertical axis of polarization of the down-converted photons. This element works as a control to turn ``on'' or ``off'' the HOM interference in the experiments described below. Then, the down-converted photons are focused into the $1cm$ quartz-cuvette containing the sample ($S$) with a $5cm$ focal length lens ($L2$), producing a $W_{0}=58\mu m$ spot diameter which is then collimated with a second lens of the same focal length ($L3$). A Rayleigh length of $Z_{R}=1.3cm$ is generated, so the interaction volume of the photons with the sample can be considered as a cylinder of $l=1cm$ length.
	
\begin{figure}[h] 
	\centering
	\includegraphics[width=85mm,height=82mm]{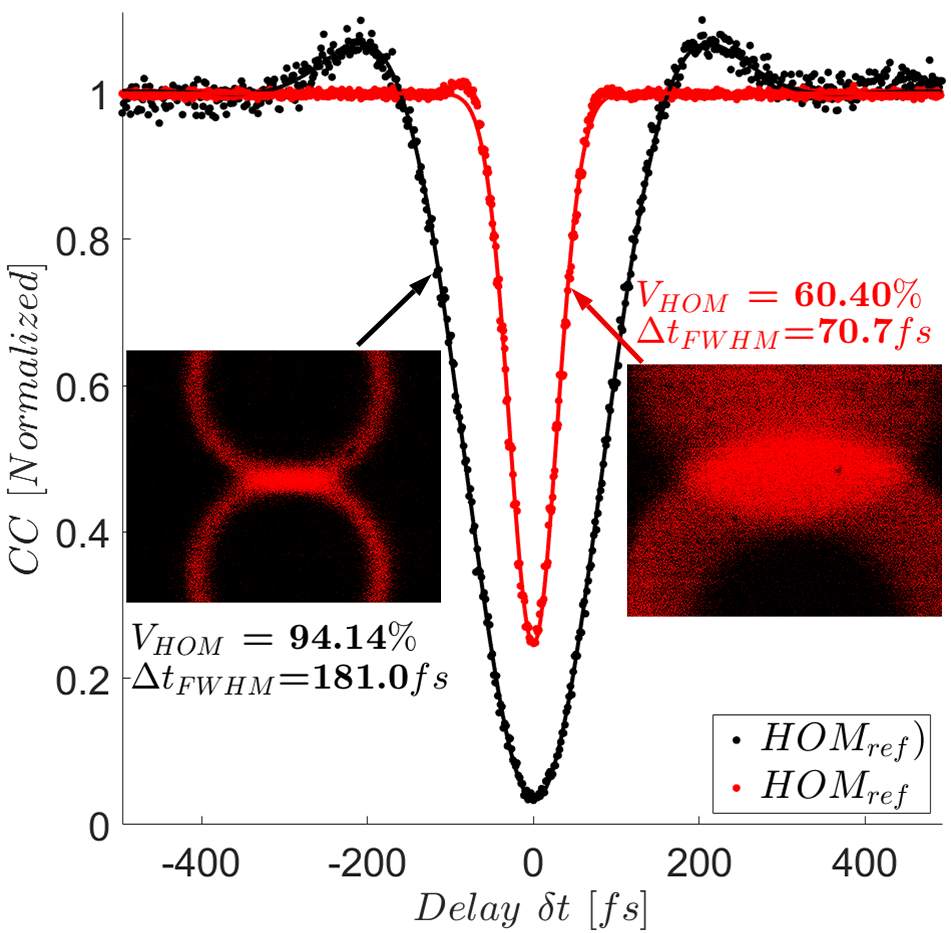}
	\caption{HOM interferogram obtained for free space propagation of the down-converted photons (cuvette removed), for two filter configurations. The black dots correspond to the HOM dip obtained with a $10$nm bandaps filter (Thorlabs FBH800-10), while the red dots where obtained with a $40$nm bandpass filter (Thorlabs FBH800-40). The insets show an image of the Type-II SPDC rings distribution as taken by a CCD camera (Thorlabs DCU224M), for both filter configurations. The collinear photons used in the experiments were obtained from the region where the SPDC cones ovelap. The solid lines were obtained from $Eq.(\ref{HOM_interferogram_2}$) by setting the ETPA sample efficiency parameter to zero ($\eta=0$), as explained below. In our experiments we use the $40$nm filters and the obtained HOM dip is denoted as $HOM_{ref}$} 
	\label{HOM_interferogram_calibration}
\end{figure}

The HOM effect results from the interference of two indistinguishable photons at a beam splitter \cite{2PI_Fearn,2PI_Pittman,2PI_Legero,SPDC_Shih,Hong_Ou_Mandel1}. In our setup we obtain the HOM interferogram by recording the coincidence counts rate $CC$ as a function of the delay time $\delta t$ between the photons, registered by a time-to-digital converter module (ID Quantique id800), between the $APD_{1}$ and $APD_{2}$ avalance photodiodes (Excelitas SPCM-AQRH), after the frequency and polarization indistinghishable SPDC down-converted photons interact at $PBS_{2}$. As a crucial property of the ETPA transmittance experiments discussed below, in our setup we can change the degree of indistinguishability of the photons in all the relevant degrees of freedom (frequency, polarization, arrival time), allowing to fully control the interference process. As shown in Fig.(\ref{HOM_interferogram_calibration}), the distinctive characteristic of the HOM interferogram is a steep dip around $\delta t = 0$ and a high coincidence counts level for $\delta t$ values larger that the coherence length of the down-converted photons.  

As an inital calibration step, we measured the HOM interferogram for free space propagation of the down-converted photons (cuvette removed), for two different configurations of the filter element $F_{1}$. The first configuration was set to optimize the spectral indistinguishability of the photon pairs and maximize the HOM dip visibility by using a bandpass filter with $10$nm bandwidth centered at $810$nm (Thorlbas FBH810-10). With this filter we obtained a HOM dip with $94\%$ visibility and a temporal FWHM of $181$fs. For the second configuration we used a bandpass filter with $40$nm bandwidth, centered at $800$nm (Thorlbas FBH800-40), obtaining a $61$\% visibility and a FWHM temporal width of $70$fs, as shown in Fig.(\ref{HOM_interferogram_calibration}). The reduction in the visibility is due to asymmetric spectrums of each photon proper of a SPDC Type-II process \cite{SPDC_Grice}. The second filter configuration will be used in the rest of the experiments and the obtained HOM dip is denoted in this work as the reference HOM interferogram $(HOM_{ref})$.

\subsection*{\label{sec:Sample-solvent-configuration}Sample and solvent configuration}
The sample used as a model in our ETPA transmittance experiments was Rhodamine B $(RhB)$ dissolved in methanol at different concentrations. As a reference, the well-known linear absorption spectrum of this molecule is displayed in Fig.(\ref{setup2}C) along with the nonlinear absorption spectrum obtained from the classical TPA effect \cite{Makarov08}. As it can be seen, around $800$nm there is region of two-photon resonance (indicated in gray color in the figure), which we will aim at in our experiments, unlike of previous work where the ETPA in $RhB$ was associated to the excitation of the state $S_{2}$ corresponding to a one-photon energy of $355$nm \cite{JuanVillabona1}. 

Two experiments were carried out to study the ETPA effect in $RhB$. In the first case we monitored the transmittance of the cuvette containing the solvent with a sample concentration of $100$mM, by detecting the coincident counts $CC$ as a function of the laser pump power, for two different $\delta t$ values $\delta t=0$ (center of HOM dip) and $\delta t=167fs$ (out of HOM dip). In the second case the laser pump power was fixed at $43.9$mW and the $CC$ were registered as a function of $\delta t$, obtaining HOM interferograms for: 1) free space propagation ($HOM_{ref}$), 2) the cuvette containing only the solvent ($HOM_{sol}$) and 3) the cuvette containing the solvent-sample solution for concentrations: $0.1\mu$M, $1\mu$M, $0.1$mM, $1$mM, $10$mM, $58$mM, $100$mM ($HOM_{sam}$). In order to avoid additional refractive errors induced by the manipulation of the cuvette, in all of these measurements the solvent-sample solutions for different concentrations were deposited without moving the cuvette, which was fixed at all time. Additionally, a cleaning and drying process was implemented every time we changed the sample under study, assuring any perturbation of the cuvette.
\begin{figure*}[htbp!]
    \centering
	\includegraphics[width=\textwidth]{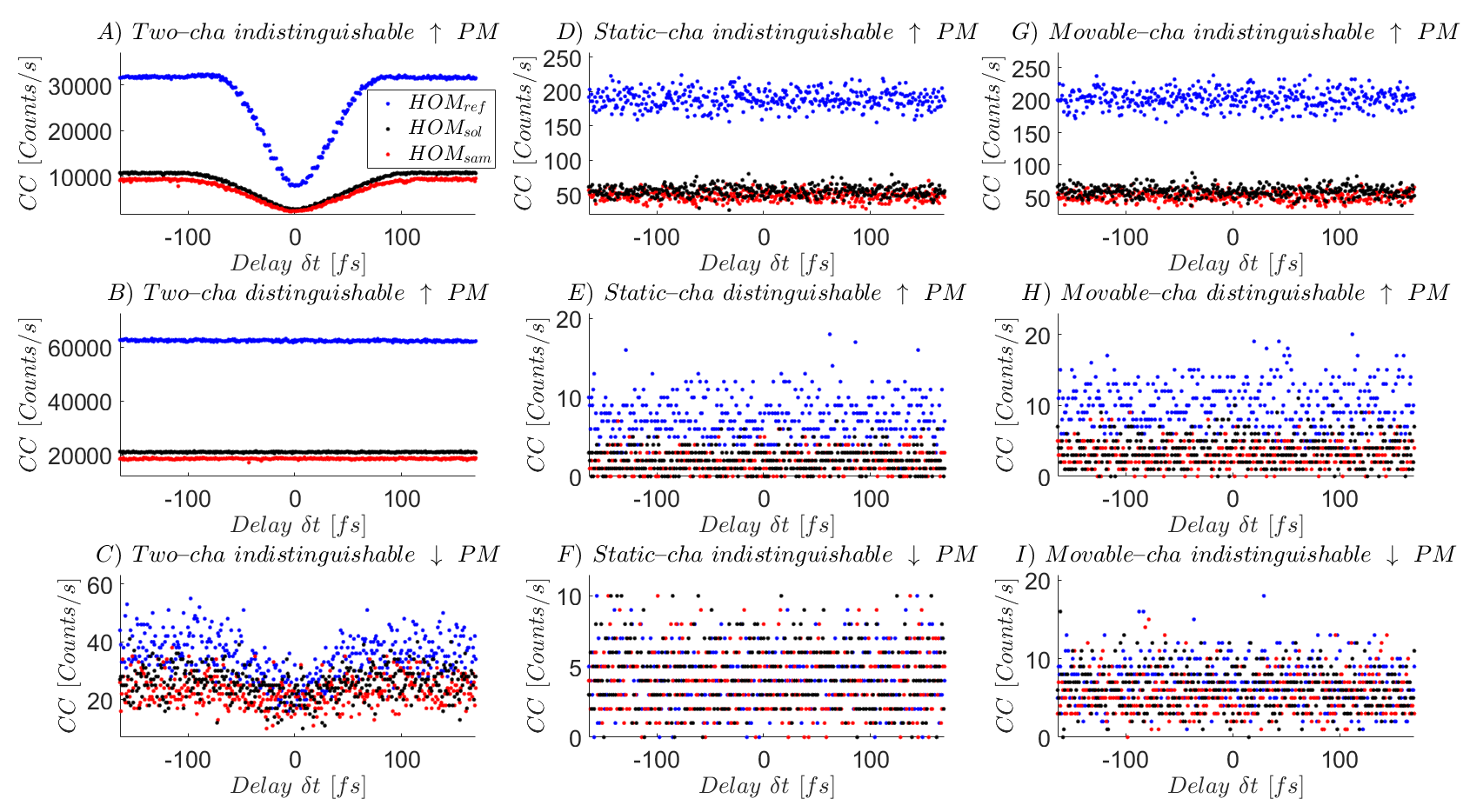}
	\caption{Noise HOM interferogram measurements. The first row present measurements taken for a configuration of indistinguishable photons and high density of down-converted photons, the second row has distinguishable photons and high density of down-converted photons and the third row has indistinguishable photons and low density of down-converted photons. $\uparrow PM$ and $\downarrow PM$ indicate high and low photon density, respectively. "$Two-cha$" indicates both arms of the interferometer unblocked, "$Static-cha$" indicates only the static arm unblocked and "$Movable-cha$" only movable arm unblocked. Red dots correspond to the measurements taken for the sample at a concentration of $100$mM ($HOM_{sam}$), black dots for the solvent ($HOM_{sol}$) and blue dots for the empty cuvette taken as the reference ($HOM_{ref}$)} 
	\label{error_HOM_interferogram}
\end{figure*}

\subsection*{\label{sec:Noise-measurement}Noise measurement}
In order to provide certainty that changes in the visibility of the HOM interferograms can be ascribed to ETPA, special care was taken to identify possible sources of noise, systematic errors and artifacts. The versatility of the experimental apparatus allows to control various parameters independently: laser pump power ($HWP_1$), density of down-converted photons ($HWMP_2$), polarization indistinguishability ($HWP_3$) and temporal delay ($\delta t$) between the SPDC photon pairs. Then, the possible presence of errors and artifacts in the HOM interferograms were assesed by monitoring the $CC$ as a function of $\delta t$, when the density of down-converted photons and the photon pairs polarization indistinguishability were set to be either high or low under the following three configurations of the HOM inteferometer: 1) both arms of the interferometer unblocked; 2) only the static arm unblocked; 3) only the delay arm unblocked. This resulted in nine different situations in which the $CC$ were plotted versus $\delta t$. In all this measurements the laser pump power was fixed constant at the maximal value of $43.9mW$. 

The first column of Fig.(\ref{error_HOM_interferogram}) presents the HOM interferograms obtained when both arms of the interferometer are unblocked ($Two-channel$). Blue dots represent the free space propagation ($HOM_{ref}$), black dots the solvent ($HOM_{sol}$) and red dots represent the sample $RhB$ in solution with methanol for a concentration of $100$mM ($HOM_{sam}$). Panel A) corresponds to the condition of polarization indistinguishable photons where, as expected, we observe a maximum HOM dip visibility, contrasting with the zero visibility produced in the case of polarization distinguishable photons (panel B)). The level of sensitivity of the apparatus can be seen in panel C) where the characteristic interference dip at $\delta t=0fs$ is obtained even under the condition of very low density of indistinguishable photons. It must be observed that for long $\delta t$, where the interference effect tend to vanish, there exist an offset between the curves in the following order: $HOM_{ref}>HOM_{sol}>HOM_{sam}$. The offset is small between sample and solvent, but significantly larger for the reference. These offsets, clearly detectable at long $\delta t$, are ascribed to linear absortion, scattering and Fresnel losses, which are larger in the sample than in solvent. Hereafter we refer to these losses as linear losses. Intriguingly, however, is the fact that the difference in $CC$ between samples and solvent decreases as the delay decreases, as observed in panel A); in particular, when the delay is zero the offset between them is minimum. It is of paramount importance to determine if such a change in the visibility of the $HOM_{sam}$ with respect to the $HOM_{sol}$ is produced by an ETPA process or it is due to artifacts in the experiment, as discussed below. 
\begin{figure*}[htbp!]
	\centering
	\includegraphics[width=\textwidth]{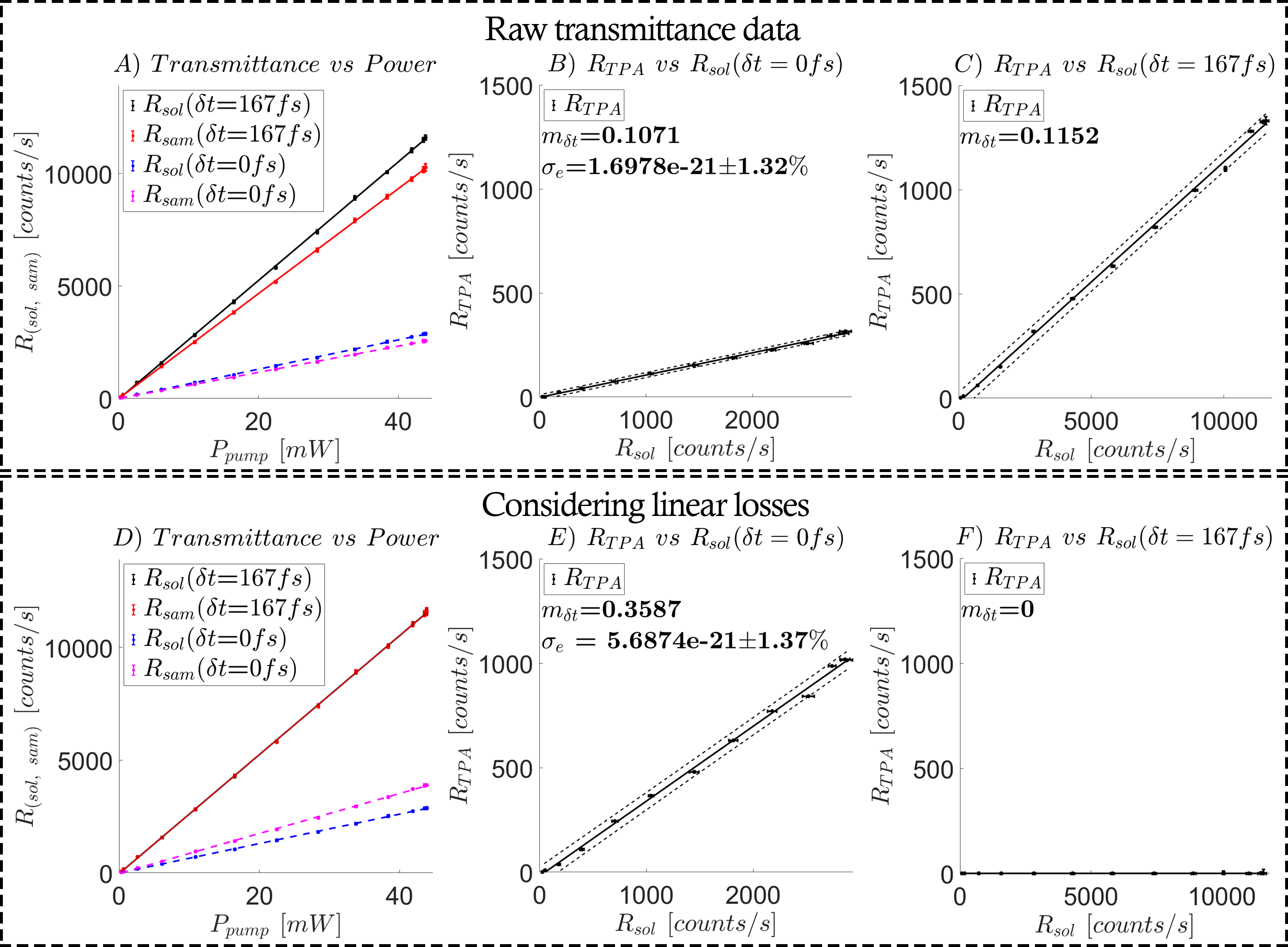}			
	\caption{ETPA transmittance measurements. Panels A) - C) present data in which linear losses are not eliminated (raw measurements), while plots in panels D) - F) show recomputed data when those linear losses are considered. A) and D) display $R_{sol}$ and $R_{sam}$ as a function of the laser pump power. B) and E) are the $R_{TPA}$ as a function of the $R_{sol}$ for a time delay of $\delta t=0$fs. Panels C) and F) are the same experiments than B) and E) but for a time delay of $\delta t=167$fs. Data in B), C), E), F) have $95\%$ predictions bounds of the fit plotted by the dashed lines.} 
	\label{CC_vs_power}
\end{figure*}
Panels D)-I) in Fig.(\ref{error_HOM_interferogram}) present the detected $CC$ when either one of the arms in the HOM interferometer is blocked. In this case the second (third) column in the figure corresponds to the detcted signal when the static (movable) arm is unblocked. As expected, the level of $CC$ is very low in these conditions as one photon of the pairs of generated entangled photons is blocked to freely propagate in the optical system, leading then to the detection of accidental $CC$. The panels D) and G) show the case of high density of indistinguishable photons, but the blocking of one of the arms leads to a level of $CC$ than can be considered basically noise. The noise is further reduced when the conditions are set to generate distinguishable photons (panels E) and H)) and becomes negligible when the density of indistinguishable photons is low (panels F) and I)). As it can be observed, in all these six cases the noise detected as accidental $CC$ is identical in each arm, demonstrating that each arm is optically equivalent. Further, it must be remarked that a dip is not registered in any of the cases in the vicinity of zero delay. Thus, the reduction of the difference in $CC$ between solvent and sample at $\delta t=0$fs, shown in panel A), should have a physical origin while it can not be explained by errors or artifacts. This suggest that the small reduction of visibility in the HOM interferograms of the sample at $\delta t=0$fs, relative to the solvent could be assigned to ETPA, being the difference of $CC$ at long delay due to a combination of absortion, scatering and diffraction losses mechanisms which do no depend on the temporal delay between the photons.

\section{\label{sec:Results}Results}
\subsection*{\label{sec:Pump-power-results}Variable pump-power measurements}
Transmittance experiments were performed to evaluate the ETPA in $RhB$. To vary the fluence of down-converted photons, the CW laser was configured to generate a pump power from $0.25mW$ to $43.9mW$. The rate of $CC$ as a function of the laser pump power was evaluated at $\delta t=0$fs and $\delta t=167$fs, for two cases: 1) cuvette with solvent and 2) cuvette with a solution of $RhB$ in methanol at a concentration of $100$mM. All these measurmentes were taken without manipulating the cuvette, which was fixed in the same position all the time, as explained before. The rate of transmitted down-converted photons detected as $CC$ for solvent ($R_{sol}$) and sample ($R_{sam}$) are presented in Fig.(\ref{CC_vs_power}A) for delay values of $\delta t=167$fs in black (solvent), red (sample), and $\delta t=0$fs in blue (solvent) and magenta (sample). The difference  $|R_{sol}-R_{sam}|$ is typically defined as the TPA rate of the sample $(R_{TPA})$ \cite{JuanVillabona1}, but special care must be taken in the use of this definition in order to discriminate the possible presence of linear losses.

Let us calculate the quotient of $R_{TPA}$ with $R_{sol}$ for the whole pump-power interval in transmittance experiments, which define the slope $m_{\delta t}=\frac{R_{TPA}}{R_{sol}}$. This slope is related with the $\sigma_{e}$ value by means of $m_{\delta t}=(CVN_{A}\sigma_{e})/(A)$  \cite{JuanVillabona1}, where $C$ is the concentration of the sample in the solvent, $V$ and $A$ are the interaction volume and area, respectively, and $N_{A}$ is the Avogadro's number. Fig.(\ref{CC_vs_power}B) presents the plot of $R_{TPA}$ versus $R_{sol}$ and the obtained fitting for $m_{\delta t}$ at $\delta t=0$fs while Fig.(\ref{CC_vs_power}C) presents the same relation for $\delta t=167$fs. As it can be seen, effectively there is a difference between $R_{sol}$ and $R_{sam}$, where $R_{TPA}$ in Fig.(\ref{CC_vs_power}C) is larger for $\delta t=167$fs than for $\delta t=0$fs in Fig.(\ref{CC_vs_power}B). However, the ETPA process should lead to a $R_{TPA}$ value larger at $\delta t=0$fs since it is a photon-delay dependent process, which takes place mainly when both photons arrive simultaneously at the sample \cite{Dayan_exp,Tabakaev}. Thus, there must be detrimental effects influencing the calculated $R_{TPA}$. 

Notice that the transmittances at the maximum power in the Fig.(\ref{CC_vs_power}A) corresponds to the condition used to generate the interferograms $HOM_{sol}$ and $HOM_{sam}$ in Fig.(\ref{error_HOM_interferogram}A) so, the difference between solvent and sample signal at $\delta t=167$fs is ascribed to linear losses which do not depend on $\delta t$. Thereby, the value of $R_{TPA}$ at $\delta t=167$fs, for each pump power value, is due exclusively to $\delta t$ independent linear losses that can be eliminated as an offset in the data of Fig.(\ref{CC_vs_power}A). When eliminating the linear losses to re-compute $R_{sam}$, the plots of panels A), B) and C) in Fig.(\ref{CC_vs_power}) produce the plots D), E) and F). In these three new graphs, various facts are noticed: first, $R_{TPA}$ for $\delta t=167$fs (panel F)) tends to be null since the down-converted photons present a high delay between them and the sample seems to be unable to produce the ETPA process. Second, the computed value of $R_{TPA}$ increase substantially for $\delta t=0$fs (panel E)) as compared with the case when the linear losses are not eliminated (panel B)). Interestingly, in figure Fig.(\ref{CC_vs_power}D) the fact that $R_{sam}$ appears larger than $R_{sol}$ for $\delta t=0$fs could be considered as a contradiction, since we expect a reduction in the transmitted photon pairs due to the ETPA interaction, but this is a direct consequence of registering the CC after the HOM interference process takes place in the sense that, as will be explained below, the ETPA interaction occurring in the sample results in a visibility decrease (CC increment) of the $HOM_{sam}$ respect to the $HOM_{sol}$, as shown in Fig.(\ref{error_HOM_interferogram}A). Third, the slope for the sample at $\delta t=0$fs is larger than at $\delta t=167$fs, meaning that if the temporal delay of the photons arriving to the sample is long, the ETPA process does not occur, conversely if the delay tends to zero, the ETPA process is detected, that is to say ETPA works as an optical switch since it can be turned "off" if the down-converted photons involved in the process arrive at the sample with a temporal delay longer than their coherence time, as measured in the HOM effect.

Various experiments on ETPA in transmission \cite{JuanVillabona1,Goodson3,JuanVillabona2} and in fluorescence\cite{Goodson1,Tabakaev} configurations have been implemented, these works have reported different values of $\sigma_{e}$ for different materials, however not considering the linear losses could have produced underestimated values of $\sigma_{e}$. In our case, when the linear losses are not taken into account a value of $\sigma_{e}=(1.6978\times10^{-21})\pm1.32\%[cm^{2}/molecule]$ is obtained while a larger value of  $\sigma_{e}=(5.6874\times10^{-21})\pm1.37\%[cm^{2}/molecule]$ is calculated by eliminating the linear losses, which is in good agreement with reported values \cite{JuanVillabona1,Samuel_Alfred}. 

\subsection*{\label{sec:HOM-results}HOM interferogram measurements results}
\begin{figure}[!t]
    \centering
	\includegraphics[width=85mm,height=82mm]{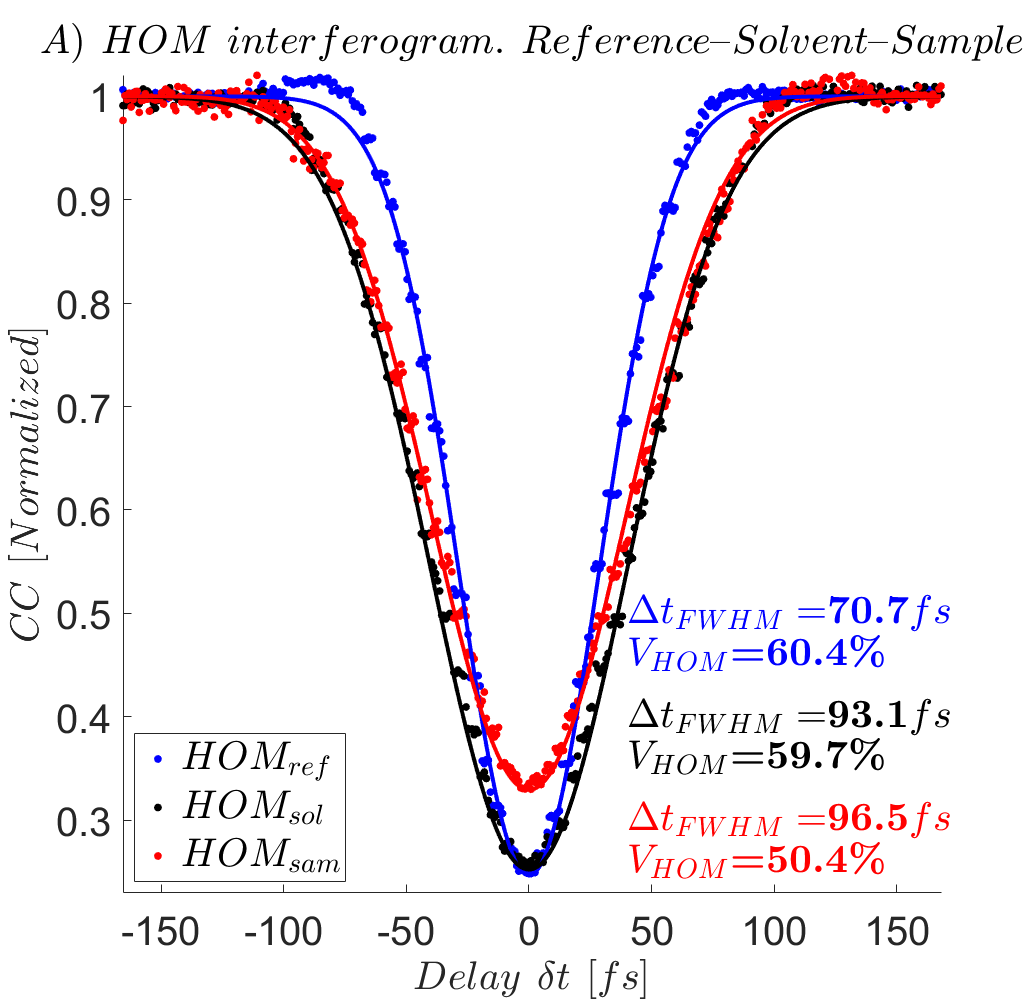}
	\includegraphics[width=85mm,height=82mm]{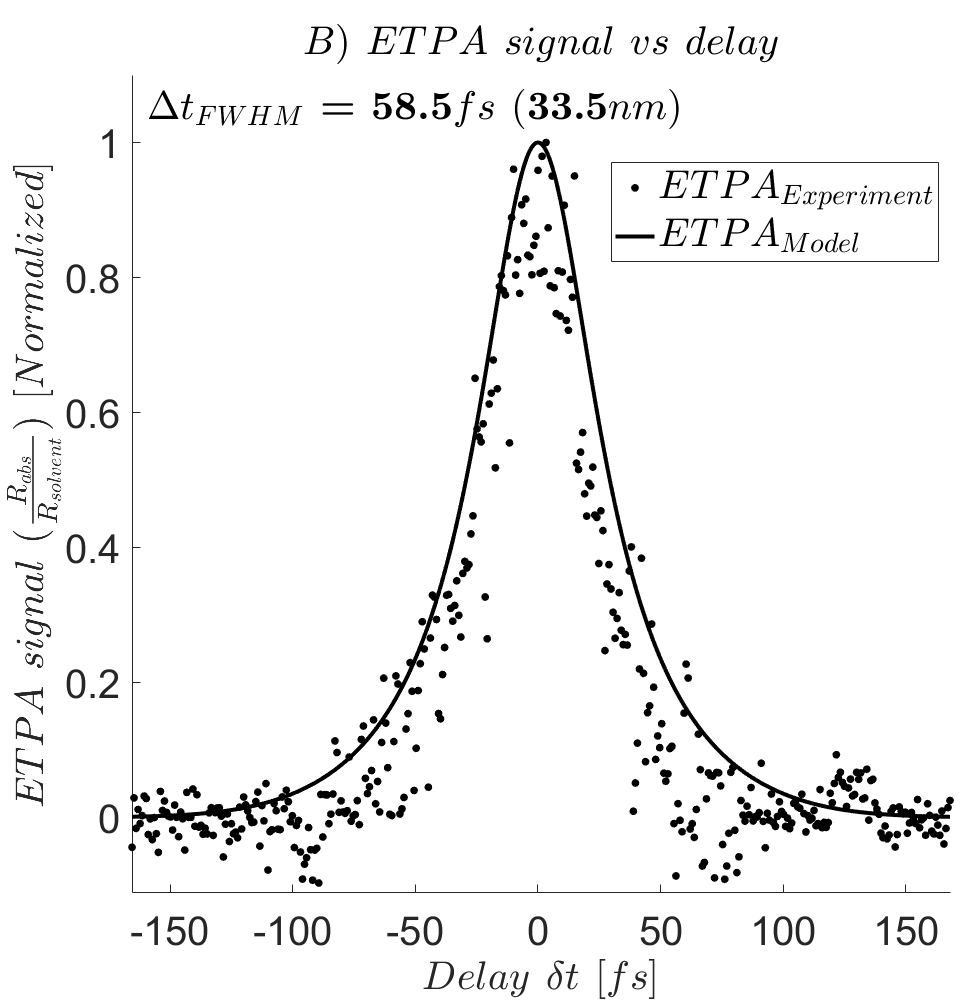}
	\caption{HOM interferogram measurements. A) Normalized HOM interferograms for sample (red line), solvent (black line) and reference (blue line). B) The ETPA signal retrieved from the solvent and sample HOM interferograms as a function of the down-converted photons delay. The solid lines on each measurement are obtained with the theoretical model presented below. Each data point of these graphs represent a measurement of 4 seconds.} 
	\label{HOM_interferogram_exp_1}
\end{figure}

Based on the results discussed above, the HOM interferograms shown in Fig.(\ref{error_HOM_interferogram}A) are revisted and presented as Fig.(\ref{HOM_interferogram_exp_1}A), eliminating the linear losses from the raw data and normalizing the interferograms. In this figure we can appreciate two main effects: first we notice that the temporal width $\Delta t_{FWHM}$ of the $HOM_{sam}$ and $HOM_{sol}$ dip increases as compared to $HOM_{ref}$. Second, it is observed that the visibility suffered a reduction for $HOM_{sam}$ and $HOM_{sol}$ as compared with $HOM_{ref}$. Both effects are present in the HOM dips of the sample and the solvent, but they are notoriously enhanced for the case of the sample. So, if ETPA did not exist, the two signals (solvent and sample) would behave similarly and the quotient between them would be a constant, but this is not the case. In Fig.(\ref{HOM_interferogram_exp_1}B) the quotient between $R_{TPA}$ and $R_{sol}$ as a function of $\delta t$ is shown. As expected, it is observed that the ETPA signal depends on the delay, showing a Gaussian-like profile with a temporal width (FWHM) of $58.5$fs.

Now, in a fundamental level, an increase in the temporal width of the HOM dip is related to a reduction of the spectral bandwidth of the interfering spectrally indistinguishable photon pairs, then the observed change in the temporal width for the $HOM_{sol}$ and $HOM_{sam}$ with respect to the $HOM_{ref}$ should be due to an absortion process which reduces the spectral bandwidth of the photons. Hence, the width increase of the solvent signal in time domain is due to a non-negligible TPA process, while in the sample the width is further increased due to a higher TPA. Then, the width change of the HOM interferogram establishes the bandwidth reduction of the down-converted photons and the visibility change establish the amount of photons absorbed (efficiency of the process), corresponding to that bandwidth reduction. In the next section we will provide a mathematical justification for the width and visibility changes of the HOM interferogram, through the modeling of the spectral interaction of the down-converted photons with the sample.
\section{\label{sec:Mathematical-model}Mathematical model}
In order to broaden the discussion of the experimental results, we propose a model based on the fundamental interaction of the down-converted photons with the two-photon transitions of the sample. With this model we find that during the ETPA interaction, the sample produces a measurable effect over the joint spectral intensity ($JSI$) function, which determines all the properties of the SPDC photon pair quantum state. Now, since the HOM dip is directly related to the JSI function of the interfering photons, the effect of the sample over the JSI can be observed as a modification in the visibility and width of the HOM interferogram, when the interference is produced after the down-converted photons interact with the sample.  
	
The state generated by a Type-II SPDC process can be written as \cite{SPDC_Adel,SPDC_Adel_2,SPDC_Sergienko}
\begin{equation}
	\lvert \psi(t)\rangle=\int_{0}^{\infty}\int_{0}^{\infty}d\omega_{s}d\omega_{i}\zeta(\omega_{s},\omega_{i})\lvert \omega_{s} \rangle_{s}\lvert \omega_{i} \rangle_{i},
	\label{estado2}
\end{equation}	
\noindent where $\lvert\ \omega_{j} \rangle$ represents a single photon with frequency $\omega_j$ for the signal (s) or idler (i) mode, with $j=s,i$. The function $\zeta(\omega_{s},\omega_{i})$ represents the joint spectral amplitud function (JSA) and $S(\omega_{s},\omega_{i})=\left\|\zeta(\omega_{s},\omega_{i})\right\|^{2}$ is the join spectral intensity (JSI). The JSA function contains all the relevant information of the photon pair quantum state and, for the case of spectrally filtered photons interacting with a sample in an ETPA process, it can be written as 
\begin{equation}
    \zeta(\omega_{s},\omega_{i})=\alpha(\omega_{s},\omega_{i})\phi(\omega_{s},\omega_{i}) f(\omega_{s})f(\omega_{i}) h(\omega_{s},\omega_{i}),
    \label{JSA}
\end{equation}
\noindent where the pump envelope function $\alpha(\omega_{s},\omega_{i})$ is given by
\begin{equation}
    \alpha(\omega_{s},\omega_{i})=e^{-\frac{\left(\omega_{s}+\omega_{i}-\omega_{p}\right)^{2}}{2\Delta\omega_{p}^{2}}},
	\label{PUMP}
\end{equation}		
\noindent with $\Delta\omega_{p}$ and $\omega_{p}$ the bandwidth and central frequency of the pump spectrum respectively, while the phase-matching function $\phi(\omega_{s},\omega_{i})$ is represented by	
\begin{equation}
	\phi(\nu_{s},\nu_{i})=e^{-\frac{\gamma}{4}\left(\tau_{s}\nu_{s}+\tau_{i}\nu_{i}\right)^{2}}=e^{-\left(P\nu_{s}^2+Q\nu_{i}^2+2R\nu_{s}\nu_{i}\right)}, 
	\label{PM}
\end{equation}		
\noindent where $\nu_{s}=\omega_{s}-\omega_{0}$, $\nu_{i}=\omega_{i}-\omega_{0}$, being $\omega_{0}=\frac{1}{2}\omega_p$ the central frequency of the down-converted photons wavepackets. Also, $P=\frac{\gamma\tau_{s}^2}{4},~~Q=\frac{\gamma\tau_{i}^2}{4},~~R=\frac{\gamma\tau_{s}\tau_{i}}{4}$ and $\gamma\simeq 0.19$ is the usual constant for the sinc to gaussian function FWHM approximation. The constants $\tau_{\mu}~(\mu=s,i)$ are related with the group velocity of the photons wavepackets by means of $\tau_{\mu}=l\left[k'_{\mu}(\omega_{0})-k'_{p}(2\omega_{0})\right]=l\left(\frac{1}{u_{\mu}}-\frac{1}{u_{p}}\right)$, being $l$ the crystal length, $k'_{\mu}$ the derivatives of the linear momentums $k_{\mu}$ of the signal and idler photons respectively and $k_p$ the linear momentum of the pump beam. Eq.(\ref{PM}) shows that the phase-matching function is not symmetric in its frequency arguments since $P\neq Q$ $(\tau_{s}\neq\tau{i})$, as a consequence of the different refractive index experimented by each photon wavepacket inside the nonlinear uniaxial crystal \cite{SPDC_Grice}.

The filter function $f(\omega_{\mu})$  accounts for the effect of the bandpass filter used to control the bandwidth of the down-converted photons ($F_1$), which has an intensity profile that can be obtained from the manufacturer datasheet and can be modeled as
\begin{equation}
	F(\omega_{\mu})=\left\| f(\omega_{\mu}) \right\|^{2}=e^{-\left[\frac{(\omega_{\mu}-\omega_{F})^{2}}{2\Delta\omega_{F}^{2}}\right]},
	\label{FILTER}
\end{equation}
\noindent where $\mu=(s,i)$ and $\omega_{F}$ and $\Delta\omega_{F}$ are the central frequency and the bandwidth of the filter respectively.

In our analysis we model the effect of the ETPA process as a filtering function $h(\omega_{s},\omega_{i})$ which fulfills the energy conservation condition required by ETPA, that is, the frequency sum of the two down-converted photons must be equal to a frequency transition of the material $\omega_{H}$, namely, $\omega_{s}+\omega_{i}=\omega_{H}$. In the frequency domain, such a sample transfer function can be modeled as a kind of "notch" filter with a Gaussian profile and bandwidth $\Delta\omega_{H}$, proposed as: 
\begin{equation}
	H(\omega_{s},\omega_{i})=\left\| h(\omega_{s},\omega_{i}) \right\|^{2}=1-\eta e^{-\frac{\left(\omega_{s}+\omega_{i}-\omega_{H}\right)^{2}}{2\Delta\omega_{H}^{2}}},
	\label{SAMPLE}
\end{equation}		
\noindent where the parameter $\eta$ ($0\leq \eta \leq 1$) quantifies the quantum efficiency of the entangled two-photon absorption process, which depends on the sample used or on the concentration of the solvent-sample solution.  
In the experiments presented here we use the two-photon interference effect as an ultra-sensitive device, capable to detect slight changes in the quantum state of the photon pairs, induced by an ETPA process occurring in the sample. Then, in order to model our measurements we need to calculate the HOM interferogram, as given by \cite{SPDC_Grice,SPDC_Timothy} 
\begin{widetext}
	\begin{equation}
	R_c(\delta t)=K\iint \limits_{-\infty}^{+\infty} \,d\omega_{s}\,d\omega_{i} \left[\left\|\Omega(\omega_{s},\omega_{i})\phi(\omega_{s},\omega_{i}) \right\|^{2} - \left\|\Omega(\omega_{s},\omega_{i}) \right\|^{2} \phi(\omega_{s},\omega_{i})\phi^{*}(\omega_{i},\omega_{s})e^{-i(\omega_{i}-\omega_{s})\delta t}\right],
    \label{HOM_interferogram}
\end{equation}
\end{widetext}
\noindent where $K$ is a normalization constant and $\Omega(\omega_{s},\omega_{i})=\alpha(\omega_{s},\omega_{i}) f(\omega_{s})f(\omega_{i})h(\omega_{s},\omega_{i})$. The first integral of Eq.(\ref{HOM_interferogram}) is the total spectral content of the down-converted photons interacting with the sample, and the second integral is the Fourier transform of the spectral product of the sample transfer function and the filtered down-converted photons function, therefore it is time-delay dependent. When the time-delay between the photons is long, the second integral of the Eq.(\ref{HOM_interferogram}) tends to zero, and $R_c(\delta t)$ will tend to the maximum level of normalized CC. Now, as the time-delay between the down converted photons decreases the subtraction of the integrals decreases until the point of zero time-delay, where $R_c(\delta t)$ will reach the minimum level of normalized CC, and it will be zero only if phase-matching function is symmetric with respect to its frequency arguments $\phi(\omega_{i},\omega_{s})=\phi(\omega_{i},\omega_{s})$. Nevertheless, if the phase-matching function is not symmetric with respect to its frequency arguments, nor if any of the functions in Eq.(\ref{HOM_interferogram}) introduces asymmetry, the subtraction of the integrals will be different from zero, and will be significantly different from zero if the asymmetry is large. The visibility of the HOM dip is given by \cite{Hong_Ou_Mandel1}:		
\begin{equation}
	V=\frac{R_{c}(\delta t)_{max}-R_{c}(\delta t)_{min}}{R_{c}(\delta t)_{max}+R_{c}(\delta t)_{min}}
	\label{VIS},
\end{equation}
\noindent where $R_{c}(\delta t)$ represent the $maximal$ and $minimal$ value of the normalized CC when scanning $\delta t$, then the visibility value will be dependent on the asymmetry of the SPDC photon pairs quantum state. 

Using the phase-matching, pump, filter and sample transfer functions described above, we can compute a final expression for the normalized HOM interferogram, which accounts for the effects of the interaction of the down-converted photons with the sample:
\begin{equation}
	R_{c}(\delta t)=1-\left[ \kappa e^{-\frac{\Delta\omega_{\Lambda}^{2}}{2}\delta t^{2}}-\eta' e^{-\frac{\Delta\omega_{J}^{2}}{2}\delta t^{2}}  \right],
    \label{HOM_interferogram_2}
\end{equation}
\noindent where the constant $\kappa$ defines the visibility of the free space HOM dip ($HOM_{ref}$), obtained when sample is not present ($\eta=0$). The function $\Delta\omega_{\Lambda}$ represents the bandwidth of the filtered down-converted photons and $\Delta\omega_{J}$ is a function which relates the bandwidths of the filtered down-converted photons function and the sample transfer function, then the width of the HOM dip will be defined and controlled by these two bandwidths. The parameter
\begin{equation}
	\eta'=\frac{J_{0}\Delta\omega_{J}}{J_{\Lambda}\Delta\omega_{\Lambda}}\eta,
    \label{eta2}
\end{equation}
\noindent is a modified ETPA quantum efficiency, which quantifies the two-photon interaction with the sample and considers the parameters:		
\begin{subequations}
	\begin{align}
		J_{0} & =e^{-\left[\frac{(\omega_{\Lambda}-\omega_{H})^{2}}{2(\Delta\omega_{\Lambda}^{2}+\Delta\omega_{H}^{2})}\right]}, \label{J0}\\
		J_{\Lambda} & =e^{-\left[\frac{(\omega_{0}-\omega_{F})^{2}}{2(\Delta\omega_{0}^{2}+\Delta\omega_{F}^{2})}\right]}, \label{JA}\\
		\Delta\omega_{J} & =\Delta\omega_{\Lambda}\left[1+\left(\frac{\Delta\omega_{\Lambda}}{\Delta\omega_{H}}\right)^{2}\right]^{-1/2},
	\end{align}
	\label{parameters_HOM}
\end{subequations}
\noindent where $\omega_{0}$ and $\Delta\omega_{0}$ are the central frequency and bandwidth of the down-converted photons and $\omega_{\Lambda}$ is the central frequency of the filtered photons. The quotient $(J_{0}/J_{\Lambda})$ presents dependence on the frequency detuning $(\delta\omega=|\omega_{\Lambda}-\omega_{H}|)$ between the central frequencies of the filtered down-converted photons and the sample transfer function. The parameter $J_{\Lambda}$ can be considered equal to one, since in general the central frequency of the filter is selected equal to the central frequency of the down-converted photons ($\omega_{F}=\omega_{0}$), thereby  $0\leq (J_{0}/J_{\Lambda}) \leq 1$. Also, in Eq.(\ref{parameters_HOM}c) $0\leq (\Delta\omega_{J}/\Delta\omega_{\Lambda}) \leq 1$ and since $0\leq \eta \leq 1$, the value of $\eta'$ must be in the range $0\leq \eta' \leq 1$. Then, the effect of the modified ETPA efficiency $\eta'$ is to reduce the visibility obtained for the reference HOM dip, as a consequence of the ETPA process. 

The model shows that there are three ways to modify the HOM interferogram related to the interaction of the down-converted photons with the sample (Eq.\ref{HOM_interferogram_2}): first, by changing the frequency detuning between the central frequencies of the filtered down-converted photons function and the sample transfer function $(\delta\omega=|\omega_{\Lambda}-\omega_{H}|)$, which has the effect to modulate $\eta'$ through $J_0$ (Eq. (\ref{eta2})), since for $\delta\omega=0$ $J_0$ is maximal and when the detuning is large $\eta' \rightarrow 0$ so we can turn ``off'' the ETPA process. Second, by varying the bandwidth of the functions $\Delta\omega_{\Lambda}$ and $\Delta\omega_{H}$ we can also modulate $\eta'$, but more importantly we can change the $\Delta\omega_J$ function which determines the temporal width of the HOM dip associated to the sample. Third, by changing the $\eta$ parameter of the sample which directly affects $\eta'$. In our experiments it is possible to modify $\eta$ by changing the concentration of the sample, while the other parameters are implicitly determined when we defined the experimental setup and when $RhB$ was chosen as a model for these studies. 
\begin{figure}[!t]
	\centering
	\includegraphics[width=85mm,height=82mm]{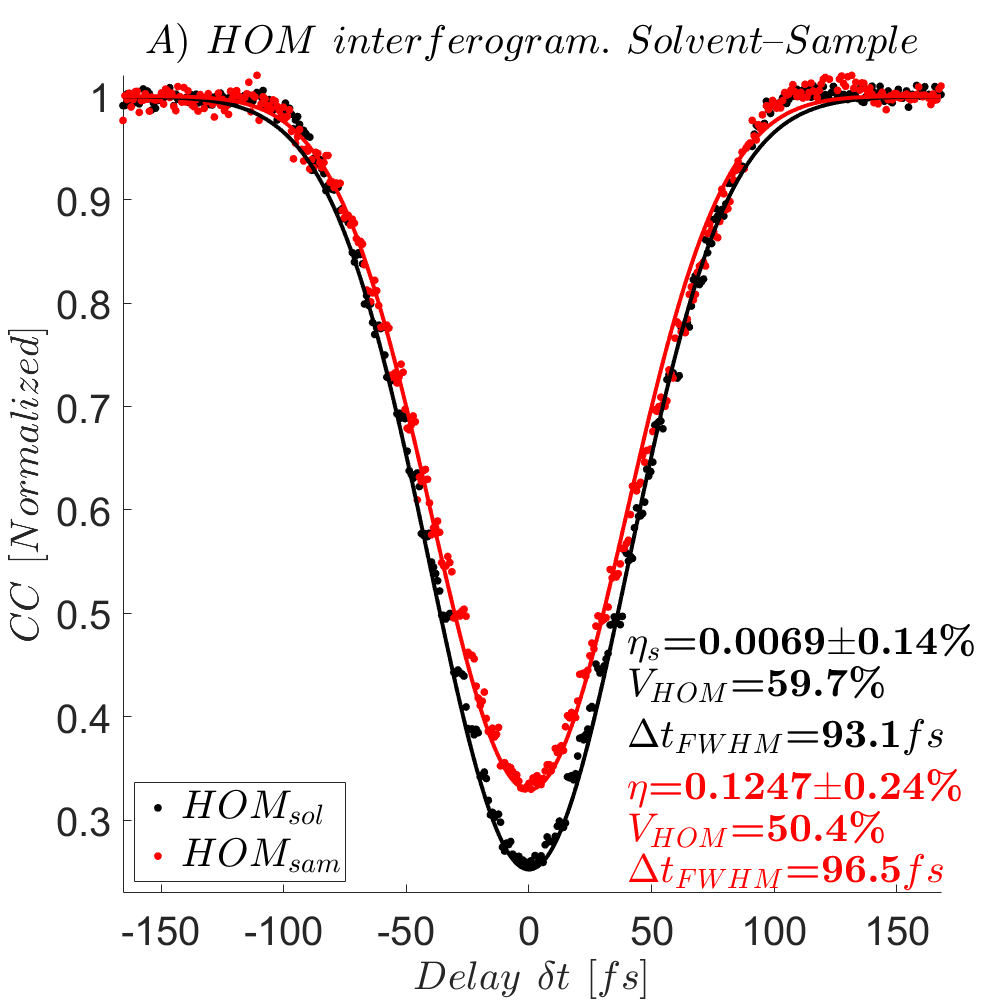}
	\includegraphics[width=85mm,height=82mm]{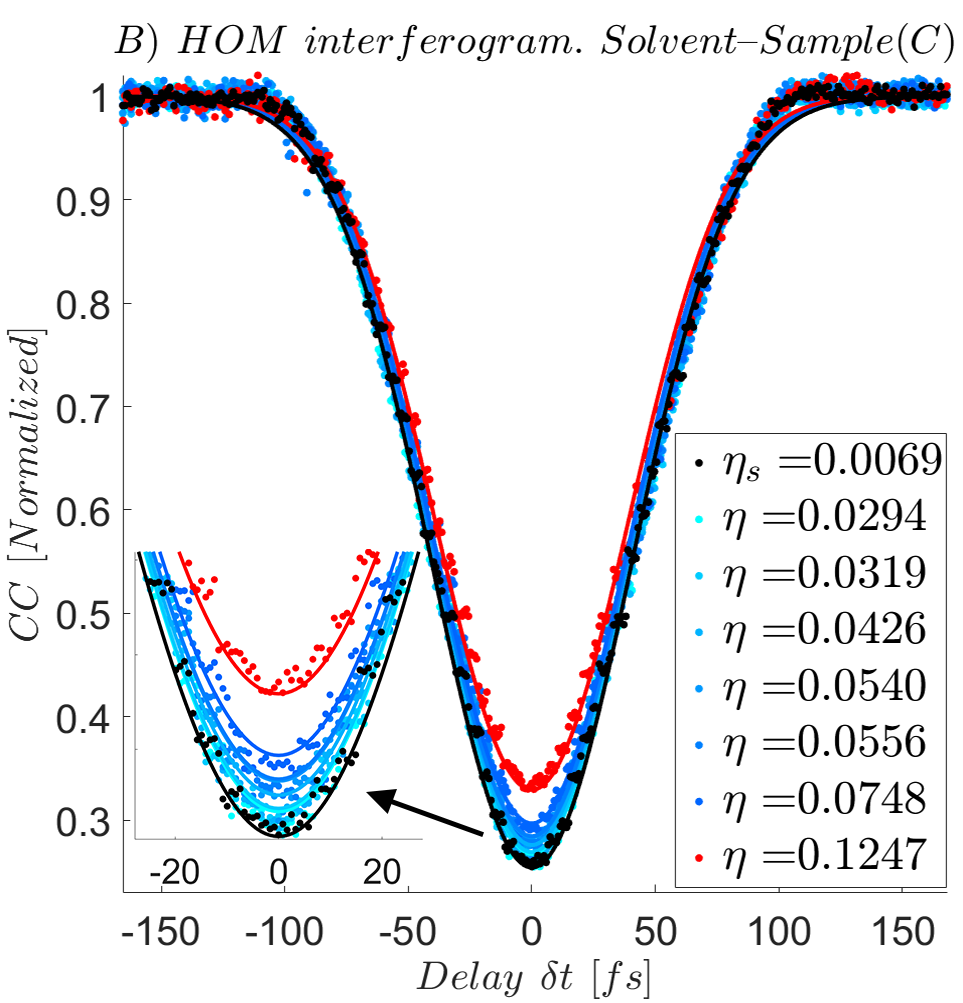}
	\caption{A) $HOM_{sam}$ for a concentration of $100$mM (red) and $HOM_{sol}$ (black). B) HOM interferogram for seven different concentrations of $RhB$ and its corresponding $\eta$ values obtained by fitting the model. The points represent the measurements and the solid lines corresponds to the theoretical curve generated by using the fitted $\eta$ value. The inset shows a zoom around the $\delta t=0$ region to appreciate how the $HOM_{sam}$ approaches to the $HOM_{sol}$ for lower concentrations. The inset also allows to observe the excellent agreement between the measurements and the theoretical model. Each data point represents a measurement of 4 seconds.} 
	\label{mathematical_model_2}
\end{figure}
To contrast our model with experimental data, we measured the HOM interferogram of the solvent-sample solution of $RhB$ for the concentrations defined in the experiment section. In total we measured eight HOM interferograms: seven for the solutions of $RhB$ ($HOM_{sam}$) and one more for the methanol solvent ($HOM_{sol}$). During these experiments we used a single 1cm long quartz-cuvette containing the solvent or the solvent-sample solutions. The cuvette was always fixed in the same position and between each successive measurement we deposited the solvent-sample solution without perturbing the cuvette, by implementing the cleaning and drying procedure described before. By fixing all the control parameters with the experimental values used in our setup, Eq.(\ref{HOM_interferogram_2}) predicts specific values of $\eta$ for each concentration of $RhB$, namely, a particular quantification of the ETPA process. 

In Fig.(\ref{mathematical_model_2}) we present the experimental results for the HOM dip measurements in dots and in solid lines the corresponding plotting of Eq.(\ref{HOM_interferogram_2}) considering the obtained ETPA process efficiency $\eta$ values. Fig.(\ref{mathematical_model_2}A) shows the HOM interferogram obtained for the solvent and for the most concentrated solution of $RhB$ ($100$mM), presenting the visibility and temporal-width values measured, along with the calculated $\eta$ values. For this sample concentration we obtain a fitted value of $\eta=0.1247$. We also present a calculated $\eta_s=0.0069$ associated with a weak ETPA process occurring in the solvent. All of the measurements and simulations, with their corresponding $\eta$ values, are presented in (\ref{mathematical_model_2}B), showing and excellent agreement as can be appreciated in the inset. In this figure we notice how for low concentrations the HOM dips obtained for the solvent and the solvent-sample solutions are almost the same, with and approaching measured visibility of $V=57.8\%$ and a corresponding predicted value of $\eta=0.0294$, for the solution at $0.1\mu$M. We can also appreciate how by increasing the concentration, or the value of the $\eta$ parameter, the HOM dip visibility correspondingly reduces, as predicted by the model. Remarkably, the model also allows to reproduce the measured ETPA signal as a function of the temporal delay between the down converted photons, presented as the solid line in Fig. (\ref{HOM_interferogram_exp_1}B). 

\section{\label{sec:Conclusion}Conclusions}
In this paper we have used the two-photon interference effect, as a mechanism to study and quantify the ETPA process for the molecular system Rhodamine B in the $800nm$ region, being the first time, to the best of our knowledge, that this approach is explored. In our setup, we measure the HOM interferogram produced by indistinguishable Type-II collinear SPDC photon pairs for three cases: 1) free space propagation, 2) after interacting with the methanol solvent, 3) after interacting with the solvent-sample solution at different concentrations. The HOM dip obtained after free space propagation is used for device optimization while the measurements obtained after the down-converted photons interact with the solvent or the solvent-sample solution are used to study the ETPA process in Rhodamine B.     

As a calibration stage we present a detailed experimental analysis of the source of errors in the coincidence counts rate, as a function of the time-delay between the down-converted photons. By direct comparison between the HOM interferograms obtained for the solvent and the solvent-sample solution, for zero temporal delay and a long temporal delay (longer than the coherence time of the down-converted photons), we can deduce the linear losses that must be compensated in order to avoid underestimations in the calculations of the ETPA cross-section. In our case, when the linear losses are not taken into account a value of $\sigma_{e}=(1.6978\times10^{-21})\pm1.32\%[cm^{2}/molecule]$ is obtained, while a larger value of  $\sigma_{e}=(5.6874\times10^{-21})\pm1.37\%[cm^{2}/molecule]$ is calculated by eliminating the linear losses. 

Then, by implementing the standard measurements of transmittance as a function of the pump power, we obtained clear evidence of the dependence of the TPA rate of the sample with the temporal delay, when considering linear losses, showing the expected behaviour of switching "on" for zero delay times and switching "off" for long delay times. Furthermore, when comparing the HOM interferogram measurements of the solvent and the solvent-sample solution it was possible to extract the ETPA signal behavior as a function of the temporal delay between the down-converted photon pairs which, also as expected, shows a clear signal peak around zero temporal delay between the indistinguishable photon pairs. This delay dependence of the ETPA signal has only been obtained in fluorescence experiments, but never from a transmission measurement.
	
We also present a detailed mathematical model which accounts for the ETPA interaction of the down-converted photons with a sample in a certain concentration, by modeling the sample as filtering function which fulfills the energy conservation conditions required by ETPA. In the model we propose the $\eta$ parameter with values $0\leq \eta \leq 1$, which accounts for the quantum efficiency of the ETPA process and, interestingly, it can be experimentally controlled by changing the concentration of the solvent-sample solution under study. We also introduce the $\Delta\omega_J$ bandwidth function which relates the bandwidths of the filtered down-converted photons function and the sample transfer function. From the model we can see that the $\Delta\omega_J$ function, fixed when the experimental setup was defined and when RhB was chosen, determines the measured width for the HOM dip associated with the ETPA process, while the $\eta$ parameter has a direct effect on the HOM dip visibility in the sense that an increasing value for $\eta$ reduces controllably the visibility, starting from a maximal value of the visibility for the HOM interferogram of the solvent. 

Then, we performed HOM dip measurements for the solvent and seven different concentrations of the solvent-sample solution and fitted the model in order to obtain predicted $\eta$ values, showing excellent agreement with the measurements. This analysis shows that for low concentration of the solvent-sample solution at $0.1\mu$M, the HOM interferograms are almost overlapped, with approaching measured visibilities of $V=57.8\%$ for the solvent-sample solution and $V=59.7\%$ for the solvent and corresponding predicted values of $\eta=0.0294$ for the sample and $\eta_s=0.0069$ for the solvent, which is considered to present a weak ETPA process. For the highest concentrated solvent-sample solution of $100m$M the measured visibility of the HOM dip reduces to $50.4\%$ with a predicted greater value of $\eta=0.1247$. It is important to mention that in the presented analysis it is missing the relation between $\eta$ and fundamental parameters of the molecular system under study, such as the $\sigma_e$ or the concentration. To find this fundamental relation will be part of another work currently under development. 

Considering this, we believe that our experimental and theoretical results represent a step forward in the application of quantum sensing techniques as ultra-sensitive devices for the study of elusive nonlinear optical effects, such as ETPA in molecular systems. 

\section{Funding}
This work was supported by CONACYT, Mexico grant Fronteras de la Ciencia 217559.

\section{Acknowledgments}
We acknowledge support from CONAYCT, Mexico.

\section{Disclosures}
The authors declare no conflicts of interest.

\end{document}